\documentclass[manuscript,screen]{acmart}
\usepackage{algorithm,algorithmic}
\usepackage{mathtools}
\DeclarePairedDelimiter\ceil{\lceil}{\rceil}
\AtBeginDocument{%
  \providecommand\BibTeX{{%
    \normalfont B\kern-0.5em{\scshape i\kern-0.25em b}\kern-0.8em\TeX}}}

\setcopyright{acmcopyright}
\copyrightyear{2022}
\acmYear{2022}
\acmDOI{XXXXXXX.XXXXXXX}

\acmConference[JEA]{ACM Journal of Experimental Algorithmics}{Feb. 20,
  2022}{Woodstock, NY}
\acmPrice{15.00}
\acmISBN{978-1-4503-XXXX-X/18/06}

\begin{document}

\title{ThielSort: Implementing the Diverting Fast Radix Algorithm}

\author{Stuart Thiel}
\email{stuart.thiel@concordia.ca}
\orcid{0000-0002-2172-3916}
\affiliation{%
  \institution{Concordia University}
  \streetaddress{1455 De Maisonneuve Blvd. W.}
  \city{Montreal}
  \state{Quebec}
  \country{Canada}
  \postcode{H3G 1M8}
}

\author{Larry Thiel}
\email{lhthiel@gmail.com}
\affiliation{%
  \institution{Concordia University}
  \streetaddress{1455 De Maisonneuve Blvd. W.}
  \city{Montreal}
  \state{Quebec}
  \country{Canada}
  \postcode{H3G 1M8}
}

\author{Gregory Butler}
\email{gregory.butler@concordia.ca}
\orcid{0000-0002-6938-0879}
\affiliation{%
  \institution{Concordia University}
  \streetaddress{1455 De Maisonneuve Blvd. W.}
  \city{Montreal}
  \state{Quebec}
  \country{Canada}
  \postcode{H3G 1M8}
}


\begin{abstract}
  This paper presents ThielSort, a practical implementation of the Diverting 
  Fast Radix (DFR) Algorithm. The theoretical improvements over classical radix 
  sorts are outlined and implementation details are specified to demonstrate
  that the algorithm is competitive with the state of the art. The 
  effectiveness of this implementation of the DFR algorithm is shown by
  considering a variety of standard distributions of data and input sizes.
\end{abstract}



\maketitle

\section{Introduction}
Exactly how Radix Sort came to be is unclear. \citeauthor{knuth1998artV3} cite's Hollerith's sorting and tabulating machines that were used in the US Census Office of the late 1800s and into the early 1900s\cite[p.384]{knuth1998artV3} as a solid early example of sorting, but \citeauthor{Cormen_2009_IntroductionToAlgorithms} summarizes it well with the following\cite[p.211]{Cormen_2009_IntroductionToAlgorithms}:

\begin{quote}
Knuth credits H.H. Seward with inventing counting sort in 1954, as well as with the idea of combining counting sort with radix sort. Radix sorting starting with the least significant digit appears to be a folk algorithm widely used by operators of mechanical card-sorting machines.
\end{quote}

Its origins as a folk algorithm, often referenced to the 1880s in terms of mechanical use, the fact that there are both most and least significant flavors of the algorithm and even changes in use of language over time lead to some confusion. In \cite{Friend_1956}, care is taken to distinguish the term \textit{digit} used in the ``Internal Digital Sorting'' as separate from the \textit{radix}, which need not be 10-based, but they note that it is ``...important to recognize that the positions in the control field must be considered in order from "least significant" to "most significant" with each one requring[sic] a separate pass''\cite[p.18]{Friend_1956}. In \cite{Rahman_Raman_2001_AdaptingRadixSortToTheMemoryHierarchy} the standard 8-bit radix is used and the terms least/most-significant-bit (LSB/MSB) first radix sort are considered, which appears to be easily interpreted as least or most-significant-byte, so that form occurs regularly. After 2000, least/most-significant-digit (LSD/MSD) radix sort, the terms we use in this paper, becomes more common in spite of \citeauthor{Friend_1956}'s careful admonishment on the use of language; in particular \citeauthor[p.211]{Cormen_2009_IntroductionToAlgorithms}'s analysis of the linear time of LSD radix sort (though they do not call it exactly that) refers to ``$d$-digit numbers in which each digit can take on up to $k$ possible values.''

There is occasional conflation with counting sort, though when only one pass is involved, they \textit{are} the same. Bucket sort is occasionally conflated with MSD radix sort. In some works, for example \cite{Lee_2002_Partitioned}, one sees the terms ``left-to-right'' or ``right-to-left'' sort that appear to be based on a big-endian interpretation of the actual encoding. \cite{Maus_2002_ARL} refers to MSD radix sort as also being ``top down radix sort'', with the ``left radix'' (again, big-endian) variety being a ``bottom up radix sort.'' \cite{Lee_2002_Partitioned} and \cite{Al-Badarneh_2004_Efficient} both refer to LSD radix sort as a ``straight radix sort''. \cite{Lorin_1975_SortingAndSortSystemes} and \cite{Andersson_Nilsson_1994_ANewEfficientRadixSort} use the term ``lexographic sort,'' though this is implied as being for variable key sized inputs (e.g. variable length strings), and it is generally of the MSD variety, as \cite{Andersson_Nilsson_1994_ANewEfficientRadixSort} also refer to it as ``forward radix sort'' and note that it is often best to divert to comparison-based-sorts when buckets become small.

The terminology used for the components of a radix sort is more consistent, with most variance being early on or based on the appearance of terms when they are considered important, and their absence otherwise. \textit{Pass} is used consistently going as far back as \cite{Friend_1956}. \cite{Friend_1956} refers to the ``recipient area'' for a particular position, but this is synonymous to what is called the \textit{buffer} in most latter papers; they do note that the ``sending area'' (what is now referred to as \textit{input}, in general) and the ``recipient area'' alternate with each pass in out-of-place implementations, which are the dominant variety of radix sort. \cite{Cormen_2009_IntroductionToAlgorithms} refers to bins. The majority of other papers refer to the specific target area for elements with like digits as \textit{buckets}.

\textit{Counting}, or frequency counting is quite consistent. Where terminology around counting varies is in consideration of performance impacts of those counts, so papers that concern themselves with these issues include mention of more details related to the counting, such as the initialization of the counts and the ``prefix sum'' of the counts\cite{Rahman_Raman_2001_AdaptingRadixSortToTheMemoryHierarchy}, and not just the count itself, which is their primary focus. 

The term for \textit{dealing} varies significantly across papers. \cite[p.18]{Friend_1956} uses the term disbursement. \cite{Rahman_Raman_2001_AdaptingRadixSortToTheMemoryHierarchy} calls this a permutation phase, what appears to be a nod to in-place MSD radix sort's unstable approach to dealing. There are also some papers that refer to the dealing phase as the distribution phase, as radix sorts fall into the class of distribution sorts; \cite{Cormen_2009_IntroductionToAlgorithms} describes how mechanical card-sorting machines would ``... examine a given column of each card in a deck and distribute the card...''\cite[p.197]{Cormen_2009_IntroductionToAlgorithms} though they do not use the term \textit{deal} that one might readily associate with \textit{deck} and \textit{card}. It is difficult to say if the term \textit{deal} is the most common term in the literature, but it is in the running.

\begin{table}
\caption[Summary of Research on Radix Sort]{This table identifies literature where improvements have been made or discussed, organized by parts of the radix sort where improvements have been applied and specific areas of research that affect the radix sort as a whole. Whether the improvement was specific to LSD or MSD radix sorts was specified for clarity.}%
\label{tab:RadixSortHistory}
\begin{tabular}{|l|l|}
\toprule
\textbf{Topic}  & \textbf{Source} \\ 
\midrule
Count & \cite{Friend_1956}, \cite{Rahman_Raman_2001_AdaptingRadixSortToTheMemoryHierarchy}, \cite{ThielImprovingGraphChi2016}, \cite{Thiel_2019}, \cite{Kumar_2019_ModifiedCountingSort}, \cite{Hanel_2020_Vortex} \\ 

Deal & \textbf{LSD:} \cite{LaMarca_1998}, \cite{Rahman_Raman_2001_AdaptingRadixSortToTheMemoryHierarchy}, \cite{Satish_2009DesigningEfficientSortingAlgorithmsForManycoreGPUs}, \cite{adinets2022onesweep} \textbf{MSD:} \cite{Jimenez-Gonzalez_Navarro_Larriba-Pey_2003_CC-Radix}, \cite{Kokot_2017_SortingDataOnUltraLargeScaleWithRADULS}, \cite{Kokot2018}, \cite{Hanel_2020_Vortex} \\ 

Diversion & \textbf{LSD:} \cite{Sedgewick_1998_AlgorithmsInCpp}, \cite{Al-Badarneh_2004_Efficient}, \cite{Thiel_2019} \textbf{MSD:} \cite{Maus_2002_ARL}, \cite{Jimenez-Gonzalez_Navarro_Larriba-Pey_2003_CC-Radix}, \cite{Kokot_2017_SortingDataOnUltraLargeScaleWithRADULS}, \cite{Kokot2018}\\ 

Parallelization & \cite{Lee_2002_Partitioned}, \cite{Satish_2009DesigningEfficientSortingAlgorithmsForManycoreGPUs}, \cite{Delorme_2013_Parallel}, \cite{adinets2022onesweep} \\ 

Variable Radix Size & \cite{LaMarca_1998}, \cite{Maus_2002_ARL} \\ 

Variable Key Size & \cite{Paige_Tarjan_1987_ThreePartitionRefinementAlgorithms}, \cite{McllroyP_Bostic_McllroyMD_1993_EngineeringRadixSort}, \cite{Andersson_Nilsson_1994_ANewEfficientRadixSort}, \cite{Bentley_Sedgewick_1997_FastAlgorithmsForSortingAndSearchingStrings}, \cite{Nilsson_2000_TheFastest}, \cite{Karkkainen_Rantala_2009_EngineeringRadixSortForStrings} \\ 
\bottomrule
\end{tabular}
\end{table}

\section{Preliminaries}
A key subset of sorting algorithms are those that sort based on the radix of data instead of by using comparisons. Each element in an input has a key that is considered in terms of a radix to break it into digits. These sorts achieve linear asymptotic complexity in all cases, and are thus generally more favorable when dealing with data that can be broken conveniently into digits.

The four common terms used by this approach are \textit{pass}, \textit{bucket}, \textit{deal} and \textit{count}. We \textit{pass} through our input. During a \textit{pass} we process each element, and this process is either a \textit{count} or a \textit{deal}, with both the dealing and the counting based on a specific digit of the key. During a single \textit{pass} we only \textit{deal} based on one digit. The \textit{count} gathers information about the occurrence of a digit in the data in order to build \textit{buckets} that we will \textit{deal} into during a subsequent \textit{pass}. We \textit{deal} records into \textit{buckets}, which means somehow moving them into a different location. Our initial input can be considered a single \textit{bucket}, but after the first \textit{pass} where we \textit{deal}, the input is organized into \textit{buckets}, and we will process the records in subsequent \textit{passes} based on which \textit{buckets} they are in.

A simple form of radix sort is a counting sort, sometimes referred to as a bucket sort, though Knuth does not make the distinction between radix and bucket (or \textit{digital}) sorts \citep[p.169]{knuth1998artV3}. In this algorithm, we consider that there is a single digit and we can pass through the input twice, counting the occurrence of each key in the first pass, and dealing into buckets during the second pass. In between the two passes, we convert the count into indices where records should be placed. With the indices established, a second pass through the input places everything in order. The areas where records will be placed, where the indices point, are referred to as buckets.

Consider an input whose contents were $N$ records with random keys between $0$ and $9$. To put this in order one could create an array $count[10]$, initialize its contents to zero, then examine each element in $N$, increasing the count $count[N[i].key]++$ for each element in $N$. This set of counts could then be converted into a set of indices where records with the same keys would be directed to the same index. The $0^{th}$ index would point to enough space to hold all the records that have a $0$ for the given digit, the next index would point to enough space to hold all the records that have a $1$ and so forth. Passing through the input once more, each record would be placed into the space allocated based on the counts, incrementing the corresponding index so it is pointing to the next free space $buff[index[N[i].key]++] = N$. When complete, the records would be sorted based on their keys, and each index would point to the space immediately following the records that had been placed.

\subsection{Least Significant Digit (LSD) Radix Sorts}
\begin{figure}
\begin{algorithm}[H]
\caption{Least Significant Digit (LSD) Radix Sort}
\label{alg:LSD}
\begin{algorithmic}[1]
\input{code/LSD.tex}
\end{algorithmic}
\end{algorithm}
\end{figure}
LSD sorts stably process data from the least significant digit to the most significant digit, performing fewer additional operations per input value, relative to a Most Significant Digit (MSD) approach that does not divert. This is how most people sort a deck of playing cards, dealing into piles with like face values before stacking those up in order and dealing into piles of like suit.

Historically, improvements to LSD radix sorts focus on improving cache-locality and skipping unneeded passes. Cache-locality improvements are generally architecture dependent, whereas pass-skipping is dependent on data distribution. Modern work on LSD radix sorts, particularly out of NVIDA, focuses on the application of LSD radix sorts parralleized across GPUs \cite{Satish_2009DesigningEfficientSortingAlgorithmsForManycoreGPUs, adinets2022onesweep}.

The pseudocode in Algorithm \ref{alg:LSD} shows a traditional approach to LSD radix sort. An initial counting pass is performed (lines \ref{alg:LSD:STARTINITIAL} - \ref{alg:LSD:ENDINITIAL}). For each digit but the last, each record is dealt into its identified space in the buffer, counting the subsequent digit and prepare the next round of counts (lines \ref{alg:LSD:STARTDEAL} - \ref{alg:LSD:ENDDEAL}). The final digit is dealt back into the original space (line \ref{alg:LSD:FINALDEAL}).

\subsection{Most Significant Digit (MSD) Radix Sorts}
MSD radix sorts process each digit, starting at the most significant digit. For each digit processed, each record will be passed over twice. In the first pass, the digits being considered are counted in order to create buckets of the appropriate size for dealing. In the second pass, the digit being considered is used to identify the properly sized bucket to deal the record into. After all digits are dealt, each bucket is sorted recursively. After each digit is processed anything in a bucket will be greater than anything in all buckets to its left and less than anything in all buckets to its right. By the time the least significant digit is processed, each bucket for that digit will either be empty or will have one or more records with the same key, and thus is entirely sorted.

A common variant on MSD radix sort is to apply diversion, what we describe as a diverting MSD radix sort. Unless all records are dealt into the same bucket, when each bucket is processed recursively it will be smaller than the original input. Processing smaller buckets is more cache efficient as buckets fit into ever more local levels of cache. Further, if buckets become small enough, diverting to another algorithm can be efficient, with anywhere from 10 to 50\% of total time of the sort involving the sorting of these tiny buckets \citet{Kokot2018}. Adinets and Merrill also identify that diversion allows parallel variants of MSD radix sort to perform significantly fewer passes as opposed to their proposed LSD algorithm\cite{adinets2022onesweep}.

MSD radix sort can be implemented in-place, but will not be stable due to the way records are swapped around. The in-place process swaps records from the free spots in identified buckets, from left to right, into the first free spots in their destination buckets. The next record from the first free spot in the left-most bucket with free spots is then processed. The savings in space come at a potential increase in cost to access records and a guaranteed increase in swaps. Ska Sort, implemented by \citet{SkaSort_2017},  is an example of a diverting MSD radix sort that makes use of many technical optimizations and ease-of-use measures to demonstrate that this type of sort is still relevant when memory must be conserved. As the performance times of in-place implementations are generally slower than corresponding ${\Theta}(n)$ space  implementations, and we will not consider in-place algorithms any further.

\begin{figure}
\begin{algorithm}[H]
\caption{Diverting Most Significant Digit (MSD) Radix Sort}
\label{alg:dMSD}
\begin{algorithmic}[1]
\input{code/dMSD.tex}
\end{algorithmic}
\end{algorithm}
\end{figure}

MSD radix sorts can be stable when implemented using an extra buffer of the same size as the input. However, diverting using a non-stable algorithm removes the stability guarantee. \citet{Kokot2018} uses sorting networks to divert in their RADULS2 implementation, and thus give up stability for the performance gains of their diversion algorithm. Using a Merge Sort as a diversion algorithm would be stable, and would take advantage of the already existing buffer space. Using Insertion Sort would be stable and would perform very well, but only on the smallest buckets.

Algorithm \ref{alg:dMSD} shows diverting MSD radix sort pseudocode given records in $input$, a suitably sized extra buffer in $buf$ and a set of digits ordered least-to-most significant in $digits$. The pseudocode starts with a check for the input being smaller than the diversion threshold (line \ref{alg:dMSD:CheckDiversionThreshold}), in which case it diverts to Insertion Sort (line \ref{alg:dMSD:Diversion}) or some other diversion algorithm. For the current most-significant digit, the algorithm counts the different values for that digit in the input and then deals into buckets located in $buf$ (lines \ref{alg:dMSD:CountDealStart}--\ref{alg:dMSD:CountDealEnd}). The algorithm is then recursively called on each bucket, using the now available space in the original input and no longer considering the now-counted digit (lines \ref{alg:dMSD:BucketsStart}--\ref{alg:dMSD:BucketsEnd}). The recursion ends all digits have been processed (lines \ref{alg:dMSD:FinishStart}--\ref{alg:dMSD:FinishEnd}).

For simplicity sake, the pseudocode in Algorithm \ref{alg:dMSD} hides some details that are less immediately relevant to the general flow of the algorithm. The pseudocode leaves out the creation and management of buckets between the count and deal functions (lines \ref{alg:dMSD:CountDealStart}--\ref{alg:dMSD:CountDealEnd}). The pseudocode hides any fiddling with offsets which may need to be done when reusing the buffer during recursion (lines \ref{alg:dMSD:BucketsStart}--\ref{alg:dMSD:BucketsEnd}). Lastly, the pseudocode of Algorithm \ref{alg:dMSD} hides the assumption that there are an even number of digits, though an actual implementation would account for either possibility.

Historically, improvements to MSD radix sorts focus on parallelization, improving cache-locality and effective diversion algorithms, with examples of all these improvements in \citet{Kokot2018}, and an example of diversion with an in-place and broadly usable implementation in \citet{SkaSort_2017}. Parallelization and cache-locality improvements are generally architecture dependent, whereas diversion is more dependent on data distribution.


\section{Diverting Fast Radix}
In previous work, \citet{ThielImprovingGraphChi2016}, we have demonstrated the improvements of Simple Fast Radix (referred to as ``Fast Radix'' in that work) over radix sorts, showing that omitting the first counting pass leads to faster performance. Here we will show how this approach can be extended and how diversion can reliably be applied to any LSD radix sort in order to create a competitively performant algorithm.

Fast Radix performs its dealing passes using estimations instead of counting for all but the last pass, including an overflow processing step in between passes. Each pass estimates bucket sizes, deals into buckets or the overflow and then processes overflow. Subsequent passes deal from each predicted bucket and then from its overflow counterpart if necessary, such that the algorithm still performs the next deal in stable, bucket order. The algorithm counts the last digit in the penultimate pass to ensure exact buckets in the last pass, avoiding an extraneous copy pass and putting everything in order.

\begin{figure}
\begin{algorithm}[H]
\caption{Fast Radix Sort}
\label{alg:FR}
\begin{algorithmic}[1]
\input{code/FR.tex}
\end{algorithmic}
\end{algorithm}
\end{figure}

Fast Radix differs from Simple Fast Radix in that only the last pass gets exact counts, with all other passes using estimates. While Simple Fast Radix could deal overflow directly to the beginning of \texttt{input} during its initial estimation pass, as there was no opportunity for conflict, Fast Radix cannot make that assumption on subsequent passes. When overflow found on a pass exceeds the overflow space already available in \texttt{overflow}, the Fast Radix algorithm creates a new \texttt{overflow} buffer that can hold twice the amount of overflow found. The algorithm can then assign overflow buckets to the first half of \texttt{overflow} and overflow from the subsequent pass can be dealt into the second half, with the algorithm dealing any excess safely into the beginning of \texttt{input}. The algorithm avoids a conflict where dealing overflow would overwrite parts of \texttt{input} that it had not yet dealt out by first dealing overflow into an overflow space that is at least as big as the existing overflow buckets.

Diverting LSD radix sort estimates the number of most significant digits over which an LSD radix sort must run to achieve optimal diversion for the remaining partially ordered records. After an LSD radix sort runs on those high-order digits, the diverting LSD radix sort will determine if any large buckets remain. Small buckets are sorted in bulk using a simple, stable sort, most readily Insertion Sort. Any large buckets are sorted recursively with diverting LSD radix sort.

\begin{figure}
\begin{algorithm}[H]
\caption{Diverting Fast Radix (DFR) Sort}
\label{alg:dFR}
\begin{algorithmic}[1]
\input{code/dFR.tex}
\end{algorithmic}
\end{algorithm}
\end{figure}

Diverting LSD radix sort differs from diverting MSD radix sort. It estimates the number of passes required to reach optimal diversion and then performs an LSD radix sort on those passes, recursing on remaining large buckets or diverting on runs of small buckets. Conversely, diverting MSD Radix sort runs recursively on each level at an ever-increasing cost before diverting. Suppose the algorithms use the same diversion threshold. In that case, we have shown that diverting LSD radix sort can accurately estimate the average number of passes for input with random records from a uniform distribution.


\subsection{Analysis}
In considering the average number of passes performed by diverting radix sorts, we will consider diverting MSD Radix Sort and our proposed diverting LSD Radix Sort. Current research on MSD Radix Sorts consider diversion as a means to reduce cost\citep{Kokot2018}. There is limited occurrence of any diverting LSD Radix Sorts, with the summary consideration of diversion after a constant number of passes\citep[Ch.10]{Sedgewick_1998_AlgorithmsInCpp} and a more recent approach involving LSD Radix Sorting floating point numbers that proposed diverting after ``$Math.min(\log{2}{n+9},32)$"\citep{Maus_2019_RadixInsert} bits. A consideration of the model for passes performed by both diverting MSD Radix Sort and our diverting LSD Radix Sort, where diversion is based on bucket size, shows that the average number of passes performed is the same in either case.

The concept of diversion is relevant to hybrid sort algorithms which utilise multiple sort algorithms at the different passes during recursion \citep{Kokot2018}. Diversion strategies in radix sort are predicated on a threshold of a bucket size $N_{d}$. If the size is below the threshold then the recursive call to radix sort on that bucket defaults to a different sort algorithm. For our analysis, we will assume the diverting sort algorithm is non-recursive so does not involve further passes. We deal with the non-diverting case by regarding it as adopting a null diversion strategy which never diverts from the radix sort.

\begin{figure}
    \centering
    \includegraphics[width=\textwidth]{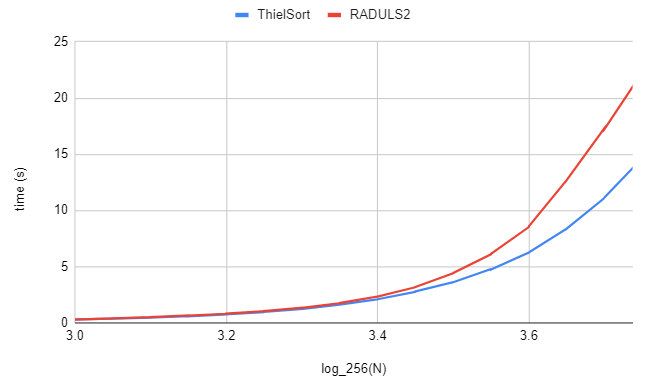}
    \caption{A comparison of ThielSort and RADULS2 on uniformly distributed integers with random inputs ranging from  $N ==256^3\sim=16$ million to $N \sim= 256^{3.75} == 1$ billion. The y-axis is time in seconds and the x-axis is $\log_{256}(N)$. At $N == 1$ billion, RADULS takes $\sim50\%$ longer.}
    \label{fig:uniform}
\end{figure}

With respect to the cost of counting, we note that traditionally it must be done for each $N$ for every pass, with the cost of managing counting in MSD radix sorts growing exponentially with $M$, a significant cost noted in \cite{adinets2022onesweep}. Conversely, estimating bucket sizes and considering overflow as a fraction of $N$, which avoids the cost of counting, can be modeled recursively as $g_{o}(N, M)$ for random inputs with uniform distributions. $N$ is the size of the input and $M$ is the number of different buckets one might deal into (the digit size).

\begin{equation}
\label{equ:overflowRecursive}
    g_o(N, M) = g_o(N-1, M) - \frac{1}{M}g_o(N-1,M,\ceil*{\frac{N}{M}}-1)
\end{equation}

Where $g_o(N, M, k)$ is defined as:
\begin{equation}
g_o(N,M,k) = (\frac{N}{M}-k)\frac{M}{N}\binom{N}{k}(\frac{1}{M})^k(1-\frac{1}{M})^{N-k}
\end{equation}

The overflow fraction $g_{o}(N, M)$ is clearly monotonically decreasing and can experimentally be shown to be a very close to zero even while $N$ is on the order of $10^3$. Those passes that use estimation, on average, save effectively the entire cost of counting, and the final pass that does count, allowing records to be placed in their sorted positions, costs no more than it would in traditional LSD or MSD radix sorts. We note also that in the event that overflow estimation is entirely wrong, it still only performs a glorified counting pass, costing comparably to traditional LSD and MSD counting passes.

While MSD radix sorts do not readily allow counting to be replaced, these results demonstrate that counting is superfluous for all passes but the last for LSD radix sorts with inputs whose least significant digits are effectively randomized according to a uniform distribution.

Considering the Marsaglia polar method (or the Knuth polar implementation) for generating values from a normal distribution, we make the conjecture that each digit of subsequently lower significance comes closer to being uniformly distributed, making estimation naturally more effective on the least significant digits even with normal distributions so long as they do not have very small standard deviations.

\begin{figure}
    \centering
    \includegraphics[width=\textwidth]{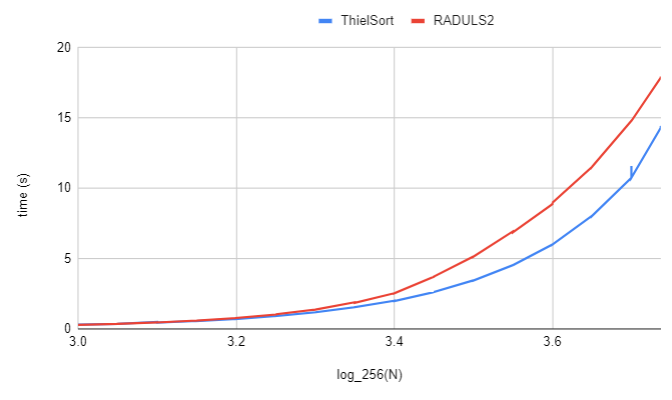}
    \caption{A comparison of ThielSort and RADULS2 on normally distributed integers with random inputs ranging from  $N ==256^3\sim=16$ million to $N \sim= 256^{3.75} == 1$ billion. The average values for the random inputs is $\sim2^{63}$ with a standard deviation of $\sim2^{61}$, what we consider a wide normal distribution. The y-axis is time in seconds and the x-axis is $\log_{256}(N)$. At $N == 1$ billion, RADULS takes $\sim25\%$ longer.}
    \label{fig:normalwide}
\end{figure}

\section{Implementation}
In implementing ThielSort, optimizations were made to render it competitive with RADULS2\cite{Kokot_2017_SortingDataOnUltraLargeScaleWithRADULS}, the current fastest radix sort. Cache-sensitive random writes, block writes and a reduction on overflow checks were the three main technical optimizations applied.

As in most current radix sort research, tackling the cost of random writes during the dealing pass of when inputs are large was required. While ThielSort does have approaches that make use of SIMD "non-temporal" writes, similarly to RADULS2\cite{Kokot_2017_SortingDataOnUltraLargeScaleWithRADULS}, in the primary range of data presented, we have found that simple pre-fetching (for writes) was sufficient to yield the desired performance gains. There are still ranges of very large ($n$ greater than 1 billion on many machines tested) where the non-temporal write strategy is more effective than pre-fetching. As well, with smaller input sizes, the cost of pre-fetching renders it slower than a vanilla deal. ThielSort's current implementation focuses on the performance in the 10s of millions to one billion range to demonstrate that with such technical optimizations DFR can outclass diverting MSD radix sorts.

In managing overflow, some dealing is performed on blocks of elements. While in uniform distributions of data and even many other distributions, very little time is spent on this process, in the cases were overflow is a more significant issue, taking advantage of faster block-writing significantly reduces the cost of these writes and modern processors have operations that support this. ThielSort makes use of this within its "ladle" system for dynamic destinations to further reduce the overhead cost of managing overflow.

A significant concern noted with DFR was that it replaced counting with an overflow check for each record on each pass. This cost can be significantly amortized across a large number of attempts by noting that if each buck starts with $\frac{N}{256}$ empty space, one can deal that many records before performing 256 bucket checks. Each subsequent round of dealing would be able to deal without checks so long as the dealing amount did not exceed the smallest bucket noted in the last round of bucket checks. When the smallest amount of space is sufficiently reduced that this trade-off is no longer advantageous, either counting every time or using some other technique can finish off. ThielSort uses this approach, using its dynamic "ladle" system and block writes when overflow becomes common. As overflow is rare, the vast majority of data is processed with a constant number of checks, rendering this cost at least constant for most data and vanishingly small for the rest for buckets with good overflow estimates. On the order of only $\frac{2N}{256^2} < 1\%$ records would remain after 256 such amortized checks if bucket estimates were accurate.

\section{Experiments} 

The experiment examines the average case performance of ThielSort, our current implementation of the Diverting Fast Radix. Results contrast ThielSort with the RADULS2 implementation of their corresponding algorithm, as it is currently the fastest published radix sort available for large n on a single processor. \texttt{std::sort} is orders of magnitude worse and not shown. 

\begin{figure}
    \centering
    \includegraphics[width=\textwidth]{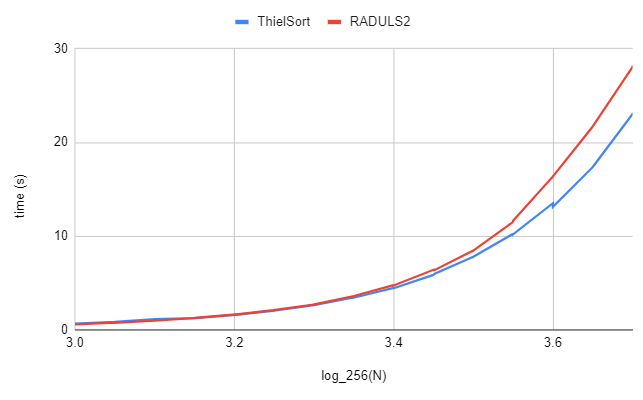}
    \caption{A comparison of ThielSort and RADULS2 on normally distributed integers with random inputs ranging from  $N ==256^3\sim=16$ million to $N \sim= 256^{3.75} == 1$ billion. The average values for the random inputs is $\sim2^{32}$ with a standard deviation of $\sim2^{30}$, what we consider a narrow normal distribution. The y-axis is time in seconds and the x-axis is $\log_{256}(N)$. At $N == 1$ billion, RADULS takes $\sim23\%$ longer.}
    \label{fig:normalnarrow}
\end{figure}

Tests were run on inputs of varying sizes from $N\sim=10^3$ to $N\sim=10^9$. Uniform distributions and a variety of normal distributions were considered, reflecting common testing approaches. Zipf distributions were also considered as \citeauthor{Kokot_2017_SortingDataOnUltraLargeScaleWithRADULS} have demonstrated their implementation's performance against those distributions, and the results reflect the narrow normal distributions shown in Figure \ref{fig:normalnarrow} while generating their input sequences much more rapidly.

While the performance reported by the RADULS2 implementation, as provided by their authors, does not include setup steps that are traditionally included in timing tests, such as allocating the extra buffer, results presented herein have included these steps in RADULS2 timings by adjusting where timings start to reflect all this work.

Tests were run on a variety of available platforms, including HPC computing systems, with Concordia's \textit{speed} HPC system being used for the test results shown here. This implementation of the Diverting Fast Radix is single-threaded and thus results are compared against RADULS2 running using a single thread. In the experimental results presented here, for each input size, three random inputs were generated and the timing results were averaged for each sort. Jobs were queues on the HPC system with sufficient ram to avoid paging to disk and each job ran alone on a single thread.

For larger input sizes, ThielSort performed significantly better in all distributions. While a wide normal distributions adversely affect ThielSort, it similarly adversely affects any other radix sort, as shown here with RADULS2. On uniform distributions, RADULS2 took $\sim50\%$ longer with inputs of size $1$ billion. On wide normal distributions, RADULS2 took $\sim25\%$ longer with inputs of size $1$ billion, which we interpret as a benefit from more early diversion, but still not making up for the extra work of an MSD radix sort. On narrow normal distributions RADULS2 took $\sim25\%$ longer with inputs of size $1$ billion, with both implementations taking longer as the distribution effectively forcing both sorts to perform more passes.

While RADULS2 still performed well on smaller input sizes, this is a function of the difference in technical optimizations used at those input-sizes; ThielSort presents a practical demonstration that the implementation of the Diverting Fast Radix can yield the predicted benefits over diverting MSD radix sorts.




\section{Conclusion and Future Research}
We presented ThielSort, an implementation of the Diverting FastRadix algorithm. The experimental results comparing ThielSort with RADULS2, the current fastest radix sort, show that the Diverting Fast Radix algorithm can be effectively implemented with technical optimizations in the same class as existing diverting MSD radix sorts to outperform them. The algorithmic improvements offered by Diverting Fast Radix are independent of hardware and exploring the affordances of the Diverting Fast Radix algorithm should allow for even more effective implementations.

Current literature highlights that current radix sort optimization focuses on parallel implementations and writing in cache efficient manners. We have already demonstrated some cache efficient approaches with this implementation, but determining optimal approaches for a given system at compile-time is an interesting avenue of future work. Parallel implementations would be advantageous, and applying previous advances in the area of LSD Radix Sorts to the Diverting Fast Radix algorithm is an obvious future target. Adinets and Merrill's recent pre-print\cite{adinets2022onesweep} highlights how actively pursued parallel LSD radix sorts are and their work represents an optimal avenue to bring ThielSort and the Diverting Fast Radix algorithm into mainstream usage.

The Diverting Fast Radix algorithm performs estimations to achieve some of its gains. Even the trivial estimations currently used offers very tangible improvements. However, approaches like sampling should be provably and efficiently able to increase performance further with non-uniform distributions. Strategies, including sampling, for dealing with smaller input sizes would make implementations of this algorithm even more broadly applicable.

Even while considering future improvements relating to implementations of our algorithm, we argue that these results already conclusively identify that LSD radix sorts are superior to MSD radix sorts when sorting data that is randomized according to a uniform distribution. Future research may support the conjecture that Diverting LSD radix sorts are superior to Diverting MSD radix sorts in all cases, i.e. there exists a corresponding LSD radix sort algorithm that provides a faster result, on average, for any MSD radix sort algorithm. We hope that this returns the attention of those researchers developing technical optimizations in this area to LSD radix sorts.

\bibliographystyle{ACM-Reference-Format}
\bibliography{References}


\begin{thebibliography}{28}


\ifx \showCODEN    \undefined \def \showCODEN     #1{\unskip}     \fi
\ifx \showDOI      \undefined \def \showDOI       #1{#1}\fi
\ifx \showISBNx    \undefined \def \showISBNx     #1{\unskip}     \fi
\ifx \showISBNxiii \undefined \def \showISBNxiii  #1{\unskip}     \fi
\ifx \showISSN     \undefined \def \showISSN      #1{\unskip}     \fi
\ifx \showLCCN     \undefined \def \showLCCN      #1{\unskip}     \fi
\ifx \shownote     \undefined \def \shownote      #1{#1}          \fi
\ifx \showarticletitle \undefined \def \showarticletitle #1{#1}   \fi
\ifx \showURL      \undefined \def \showURL       {\relax}        \fi
\providecommand\bibfield[2]{#2}
\providecommand\bibinfo[2]{#2}
\providecommand\natexlab[1]{#1}
\providecommand\showeprint[2][]{arXiv:#2}

\bibitem[Adinets and Merrill(2022)]%
        {adinets2022onesweep}
\bibfield{author}{\bibinfo{person}{Andy Adinets} {and} \bibinfo{person}{Duane
  Merrill}.} \bibinfo{year}{2022}\natexlab{}.
\newblock \showarticletitle{Onesweep: A Faster Least Significant Digit Radix
  Sort for GPUs}.
\newblock \bibinfo{journal}{\emph{arXiv preprint arXiv:2206.01784}}
  (\bibinfo{year}{2022}).
\newblock


\bibitem[Al-Badarneh and El-Aker(2004)]%
        {Al-Badarneh_2004_Efficient}
\bibfield{author}{\bibinfo{person}{Amer Al-Badarneh} {and}
  \bibinfo{person}{Fouad El-Aker}.} \bibinfo{year}{2004}\natexlab{}.
\newblock \showarticletitle{Efficient Adaptive In‐Place Radix Sorting.}
\newblock \bibinfo{journal}{\emph{Informatica}} \bibinfo{volume}{15},
  \bibinfo{number}{3} (\bibinfo{year}{2004}), \bibinfo{pages}{295 -- 302}.
\newblock
\showISSN{08684952}
\urldef\tempurl%
\url{https://lib-ezproxy.concordia.ca/login?url=https://search.ebscohost.com/login.aspx?direct=true&db=a9h&AN=18106146&site=ehost-live&scope=site}
\showURL{%
\tempurl}


\bibitem[Andersson and Nilsson(1994)]%
        {Andersson_Nilsson_1994_ANewEfficientRadixSort}
\bibfield{author}{\bibinfo{person}{Arne Andersson} {and}
  \bibinfo{person}{Stefan Nilsson}.} \bibinfo{year}{1994}\natexlab{}.
\newblock \showarticletitle{A new efficient radix sort}. In
  \bibinfo{booktitle}{\emph{Proceedings, 35th Annual Symposium on Foundations
  of Computer Science, 1994}}. \bibinfo{publisher}{IEEE}, \bibinfo{address}{New
  York, NY, USA}, \bibinfo{pages}{714--721}.
\newblock
\urldef\tempurl%
\url{https://doi.org/10.1109/SFCS.1994.365721}
\showDOI{\tempurl}


\bibitem[Bentley and Sedgewick(1997)]%
        {Bentley_Sedgewick_1997_FastAlgorithmsForSortingAndSearchingStrings}
\bibfield{author}{\bibinfo{person}{Jon~L. Bentley} {and}
  \bibinfo{person}{Robert Sedgewick}.} \bibinfo{year}{1997}\natexlab{}.
\newblock \showarticletitle{Fast Algorithms for Sorting and Searching Strings}.
  In \bibinfo{booktitle}{\emph{Proceedings of the Eighth Annual ACM-SIAM
  Symposium on Discrete Algorithms}} (New Orleans, Louisiana, USA).
  \bibinfo{publisher}{ACM}, \bibinfo{address}{Philadelphia, PA, USA},
  \bibinfo{pages}{360--369}.
\newblock
\showISBNx{0-89871-390-0}


\bibitem[Cormen et~al\mbox{.}(2009)]%
        {Cormen_2009_IntroductionToAlgorithms}
\bibfield{author}{\bibinfo{person}{Thomas~H. Cormen},
  \bibinfo{person}{Charles~E. Leiserson}, \bibinfo{person}{Ronald~L. Rivest},
  {and} \bibinfo{person}{Clifford Stein}.} \bibinfo{year}{2009}\natexlab{}.
\newblock \bibinfo{booktitle}{\emph{Introduction to Algorithms, 3rd Edition}}.
\newblock \bibinfo{publisher}{{MIT} Press}, \bibinfo{address}{Cambridge, MA,
  USA}.
\newblock
\showISBNx{978-0-262-03384-8}
\urldef\tempurl%
\url{http://mitpress.mit.edu/books/introduction-algorithms}
\showURL{%
\tempurl}


\bibitem[{Delorme} et~al\mbox{.}(2013)]%
        {Delorme_2013_Parallel}
\bibfield{author}{\bibinfo{person}{Michael~C. {Delorme}},
  \bibinfo{person}{Tarek~S. {Abdelrahman}}, {and} \bibinfo{person}{Chengyan
  {Zhao}}.} \bibinfo{year}{2013}\natexlab{}.
\newblock \showarticletitle{Parallel Radix Sort on the AMD Fusion Accelerated
  Processing Unit}. In \bibinfo{booktitle}{\emph{2013 42nd International
  Conference on Parallel Processing}}. \bibinfo{publisher}{IEEE},
  \bibinfo{address}{New York, NY, USA}, \bibinfo{pages}{339--348}.
\newblock
\urldef\tempurl%
\url{https://doi.org/10.1109/ICPP.2013.43}
\showDOI{\tempurl}


\bibitem[Friend(1956)]%
        {Friend_1956}
\bibfield{author}{\bibinfo{person}{Edward~H. Friend}.}
  \bibinfo{year}{1956}\natexlab{}.
\newblock \showarticletitle{Sorting on Electronic Computer Systems}.
\newblock \bibinfo{journal}{\emph{J. ACM}} \bibinfo{volume}{3},
  \bibinfo{number}{3} (\bibinfo{date}{July} \bibinfo{year}{1956}),
  \bibinfo{pages}{134–168}.
\newblock
\showISSN{0004-5411}
\urldef\tempurl%
\url{https://doi.org/10.1145/320831.320833}
\showDOI{\tempurl}


\bibitem[Hanel et~al\mbox{.}(2020)]%
        {Hanel_2020_Vortex}
\bibfield{author}{\bibinfo{person}{Carson Hanel}, \bibinfo{person}{Arif Arman},
  \bibinfo{person}{Di Xiao}, \bibinfo{person}{John Keech}, {and}
  \bibinfo{person}{Dmitri Loguinov}.} \bibinfo{year}{2020}\natexlab{}.
\newblock \bibinfo{booktitle}{\emph{Vortex: Extreme-Performance Memory
  Abstractions for Data-Intensive Streaming Applications}}.
\newblock \bibinfo{publisher}{Association for Computing Machinery},
  \bibinfo{address}{New York, NY, USA}, \bibinfo{pages}{623–638}.
\newblock
\showISBNx{9781450371025}
\urldef\tempurl%
\url{https://doi.org/10.1145/3373376.3378527}
\showURL{%
\tempurl}


\bibitem[Jim{\'e}nez-Gonz{\'a}lez et~al\mbox{.}(2003)]%
        {Jimenez-Gonzalez_Navarro_Larriba-Pey_2003_CC-Radix}
\bibfield{author}{\bibinfo{person}{Daniel Jim{\'e}nez-Gonz{\'a}lez},
  \bibinfo{person}{E Guinovart}, \bibinfo{person}{J-L Larriba-Pey}, {and}
  \bibinfo{person}{Juan~J Navarro}.} \bibinfo{year}{2003}\natexlab{}.
\newblock \showarticletitle{{CC-Radix}: a cache conscious sorting based on
  {Radix sort}}. In \bibinfo{booktitle}{\emph{Parallel, Distributed and
  Network-Based Processing, 2003. Proceedings. Eleventh Euromicro Conference
  on}}. \bibinfo{publisher}{IEEE}, \bibinfo{address}{New York, NY, USA},
  \bibinfo{pages}{101--108}.
\newblock
\showISSN{1066-6192}
\urldef\tempurl%
\url{https://doi.org/10.1109/EMPDP.2003.1183573}
\showDOI{\tempurl}


\bibitem[K{\"a}rkk{\"a}inen and Rantala(2009)]%
        {Karkkainen_Rantala_2009_EngineeringRadixSortForStrings}
\bibfield{author}{\bibinfo{person}{Juha K{\"a}rkk{\"a}inen} {and}
  \bibinfo{person}{Tommi Rantala}.} \bibinfo{year}{2009}\natexlab{}.
\newblock \showarticletitle{Engineering Radix Sort for Strings}. In
  \bibinfo{booktitle}{\emph{String Processing and Information Retrieval}},
  \bibfield{editor}{\bibinfo{person}{Amihood Amir}, \bibinfo{person}{Andrew
  Turpin}, {and} \bibinfo{person}{Alistair Moffat}} (Eds.).
  \bibinfo{publisher}{Springer Berlin Heidelberg}, \bibinfo{address}{Berlin,
  Heidelberg}, \bibinfo{pages}{3--14}.
\newblock
\showISBNx{978-3-540-89097-3}


\bibitem[Knuth(1998)]%
        {knuth1998artV3}
\bibfield{author}{\bibinfo{person}{Donald~E. Knuth}.}
  \bibinfo{year}{1998}\natexlab{}.
\newblock \bibinfo{booktitle}{\emph{{The Art of Computer Programming}}}.
\newblock Number v. 3 in \bibinfo{series}{Addison-Wesley series in computer
  science and information processing}. \bibinfo{publisher}{Addison-Wesley},
  \bibinfo{address}{Boston, MA, USA}.
\newblock
\showISBNx{9780201896855}
\showLCCN{97002147}
\urldef\tempurl%
\url{https://books.google.ca/books?id=CH1GAAAAYAAJ}
\showURL{%
\tempurl}


\bibitem[Kokot et~al\mbox{.}(2017)]%
        {Kokot_2017_SortingDataOnUltraLargeScaleWithRADULS}
\bibfield{author}{\bibinfo{person}{Marek Kokot}, \bibinfo{person}{Sebastian
  Deorowicz}, {and} \bibinfo{person}{Agnieszka Debudaj-Grabysz}.}
  \bibinfo{year}{2017}\natexlab{}.
\newblock \showarticletitle{Sorting Data on Ultra-Large Scale with RADULS}. In
  \bibinfo{booktitle}{\emph{Beyond Databases, Architectures and Structures.
  Towards Efficient Solutions for Data Analysis and Knowledge Representation}},
  \bibfield{editor}{\bibinfo{person}{Stanis{\l}aw Kozielski},
  \bibinfo{person}{Dariusz Mrozek}, \bibinfo{person}{Pawe{\l} Kasprowski},
  \bibinfo{person}{Bo{\.{z}}ena Ma{\l}ysiak-Mrozek}, {and}
  \bibinfo{person}{Daniel Kostrzewa}} (Eds.). \bibinfo{publisher}{Springer
  International Publishing}, \bibinfo{address}{New York, NY, USA},
  \bibinfo{pages}{235--245}.
\newblock
\showISBNx{978-3-319-58274-0}


\bibitem[Kokot et~al\mbox{.}(2018)]%
        {Kokot2018}
\bibfield{author}{\bibinfo{person}{Marek Kokot}, \bibinfo{person}{Sebastian
  Deorowicz}, {and} \bibinfo{person}{Maciej D{\l}ugosz}.}
  \bibinfo{year}{2018}\natexlab{}.
\newblock \showarticletitle{Even Faster Sorting of (Not Only) Integers}.
\newblock In \bibinfo{booktitle}{\emph{Man-Machine Interactions 5}},
  \bibfield{editor}{\bibinfo{person}{Aleksandra Gruca},
  \bibinfo{person}{Tadeusz Czach{\'o}rski}, \bibinfo{person}{Katarzyna
  Harezlak}, \bibinfo{person}{Stanis{\l}aw Kozielski}, {and}
  \bibinfo{person}{Agnieszka Piotrowska}} (Eds.).
  \bibinfo{publisher}{Springer}, \bibinfo{address}{New York, NY, USA},
  \bibinfo{pages}{481--491}.
\newblock
\urldef\tempurl%
\url{https://doi.org/10.1007/978-3-319-67792-7_47}
\showDOI{\tempurl}


\bibitem[Kumar(2019)]%
        {Kumar_2019_ModifiedCountingSort}
\bibfield{author}{\bibinfo{person}{Ravin Kumar}.}
  \bibinfo{year}{2019}\natexlab{}.
\newblock \showarticletitle{Modified Counting Sort}.
\newblock In \bibinfo{booktitle}{\emph{System Performance and Management
  Analytics}}, \bibfield{editor}{\bibinfo{person}{P.~K. Kapur},
  \bibinfo{person}{Yury Klochkov}, \bibinfo{person}{Ajit~Kumar Verma}, {and}
  \bibinfo{person}{Gurinder Singh}} (Eds.). \bibinfo{publisher}{Springer
  Singapore}, \bibinfo{address}{Singapore}, \bibinfo{pages}{251--258}.
\newblock
\showISBNx{978-981-10-7323-6}
\urldef\tempurl%
\url{https://doi.org/10.1007/978-981-10-7323-6_21}
\showDOI{\tempurl}


\bibitem[LaMarca and Ladner(1999)]%
        {LaMarca_1998}
\bibfield{author}{\bibinfo{person}{Anthony LaMarca} {and}
  \bibinfo{person}{Richard~E Ladner}.} \bibinfo{year}{1999}\natexlab{}.
\newblock \showarticletitle{The Influence of Caches on the Performance of
  Sorting}.
\newblock \bibinfo{journal}{\emph{Journal of Algorithms}} \bibinfo{volume}{31},
  \bibinfo{number}{1} (\bibinfo{year}{1999}), \bibinfo{pages}{66 -- 104}.
\newblock
\showISSN{0196-6774}
\urldef\tempurl%
\url{https://doi.org/10.1006/jagm.1998.0985}
\showDOI{\tempurl}


\bibitem[Lee et~al\mbox{.}(2002)]%
        {Lee_2002_Partitioned}
\bibfield{author}{\bibinfo{person}{Shin-Jae Lee}, \bibinfo{person}{Minsoo
  Jeon}, \bibinfo{person}{Dongseung Kim}, {and} \bibinfo{person}{Andrew Sohn}.}
  \bibinfo{year}{2002}\natexlab{}.
\newblock \showarticletitle{Partitioned Parallel Radix Sort}.
\newblock \bibinfo{journal}{\emph{J. Parallel and Distrib. Comput.}}
  \bibinfo{volume}{62}, \bibinfo{number}{4} (\bibinfo{year}{2002}),
  \bibinfo{pages}{656 -- 668}.
\newblock
\showISSN{0743-7315}
\urldef\tempurl%
\url{https://doi.org/10.1006/jpdc.2001.1808}
\showDOI{\tempurl}


\bibitem[Lorin(1975)]%
        {Lorin_1975_SortingAndSortSystemes}
\bibfield{author}{\bibinfo{person}{Harold Lorin}.}
  \bibinfo{year}{1975}\natexlab{}.
\newblock \bibinfo{booktitle}{\emph{Sorting and Sort Systems (The Systems
  Programming Series)}}.
\newblock \bibinfo{publisher}{Addison-Wesley Longman Publishing Co., Inc.},
  \bibinfo{address}{Boston, MA, USA}.
\newblock
\showISBNx{0201144530}


\bibitem[Maus(2002)]%
        {Maus_2002_ARL}
\bibfield{author}{\bibinfo{person}{Arne Maus}.}
  \bibinfo{year}{2002}\natexlab{}.
\newblock \showarticletitle{ARL, a faster in-place, cache friendly sorting
  algorithm}. In \bibinfo{booktitle}{\emph{Norsk Informatik Konferranse
  NIK’2002}}. \bibinfo{publisher}{NIKT}, \bibinfo{address}{Norway},
  \bibinfo{pages}{85--95}.
\newblock


\bibitem[Maus(2019)]%
        {Maus_2019_RadixInsert}
\bibfield{author}{\bibinfo{person}{Arne Maus}.}
  \bibinfo{year}{2019}\natexlab{}.
\newblock \showarticletitle{RadixInsert, a much faster stable algorithm for
  sorting floating-point numbers}. In \bibinfo{booktitle}{\emph{Norsk
  IKT-Konferanse for Forskning og Utdanning}}. \bibinfo{publisher}{NIKT},
  \bibinfo{address}{Norway}.
\newblock


\bibitem[Mcllroy et~al\mbox{.}(1993)]%
        {McllroyP_Bostic_McllroyMD_1993_EngineeringRadixSort}
\bibfield{author}{\bibinfo{person}{Peter~M Mcllroy}, \bibinfo{person}{Keith
  Bostic}, {and} \bibinfo{person}{Malcolm~Douglas Mcllroy}.}
  \bibinfo{year}{1993}\natexlab{}.
\newblock \showarticletitle{Engineering {Radix Sort}}.
\newblock \bibinfo{journal}{\emph{{Computing Systems}}} \bibinfo{volume}{6},
  \bibinfo{number}{1} (\bibinfo{year}{1993}), \bibinfo{pages}{5--27}.
\newblock


\bibitem[Nilsson(2000)]%
        {Nilsson_2000_TheFastest}
\bibfield{author}{\bibinfo{person}{Stefan Nilsson}.}
  \bibinfo{year}{2000}\natexlab{}.
\newblock \showarticletitle{{The Fastest Sorting Algorithm?}}
\newblock \bibinfo{journal}{\emph{Dr. Dobb's journal (1989)}}
  \bibinfo{volume}{25}, \bibinfo{number}{4} (\bibinfo{year}{2000}),
  \bibinfo{pages}{38--+}.
\newblock


\bibitem[Paige and Tarjan(1987)]%
        {Paige_Tarjan_1987_ThreePartitionRefinementAlgorithms}
\bibfield{author}{\bibinfo{person}{Robert Paige} {and}
  \bibinfo{person}{Robert~E Tarjan}.} \bibinfo{year}{1987}\natexlab{}.
\newblock \showarticletitle{Three partition refinement algorithms}.
\newblock \bibinfo{journal}{\emph{SIAM J. Comput.}} \bibinfo{volume}{16},
  \bibinfo{number}{6} (\bibinfo{year}{1987}), \bibinfo{pages}{973--989}.
\newblock


\bibitem[Rahman and Raman(2002)]%
        {Rahman_Raman_2001_AdaptingRadixSortToTheMemoryHierarchy}
\bibfield{author}{\bibinfo{person}{Naila Rahman} {and} \bibinfo{person}{Rajeev
  Raman}.} \bibinfo{year}{2002}\natexlab{}.
\newblock \showarticletitle{Adapting Radix Sort to the Memory Hierarchy}.
\newblock \bibinfo{journal}{\emph{ACM J. Exp. Algorithmics}}
  \bibinfo{volume}{6} (\bibinfo{date}{dec} \bibinfo{year}{2002}),
  \bibinfo{pages}{7–es}.
\newblock
\showISSN{1084-6654}
\urldef\tempurl%
\url{https://doi.org/10.1145/945394.945401}
\showDOI{\tempurl}


\bibitem[{Satish} et~al\mbox{.}(2009)]%
        {Satish_2009DesigningEfficientSortingAlgorithmsForManycoreGPUs}
\bibfield{author}{\bibinfo{person}{Nadathur {Satish}}, \bibinfo{person}{Mark
  {Harris}}, {and} \bibinfo{person}{Michael {Garland}}.}
  \bibinfo{year}{2009}\natexlab{}.
\newblock \showarticletitle{Designing efficient sorting algorithms for manycore
  GPUs}. In \bibinfo{booktitle}{\emph{2009 IEEE International Symposium on
  Parallel Distributed Processing}}. \bibinfo{publisher}{IEEE},
  \bibinfo{address}{New York, NY, USA}, \bibinfo{pages}{1--10}.
\newblock
\urldef\tempurl%
\url{https://doi.org/10.1109/IPDPS.2009.5161005}
\showDOI{\tempurl}


\bibitem[Sedgewick(1998)]%
        {Sedgewick_1998_AlgorithmsInCpp}
\bibfield{author}{\bibinfo{person}{Robert Sedgewick}.}
  \bibinfo{year}{1998}\natexlab{}.
\newblock \bibinfo{booktitle}{\emph{Algorithms in C++}}.
\newblock Number v. 1-2 in \bibinfo{series}{Algorithms in C++}.
  \bibinfo{publisher}{Addison-Wesley}, \bibinfo{address}{Boston, MA, USA}.
\newblock
\showISBNx{9780201350883}
\showLCCN{97023418}
\urldef\tempurl%
\url{https://books.google.ca/books?id=331YAAAAYAAJ}
\showURL{%
\tempurl}


\bibitem[Skarupke(2017)]%
        {SkaSort_2017}
\bibfield{author}{\bibinfo{person}{Malte Skarupke}.}
  \bibinfo{year}{2017}\natexlab{}.
\newblock \bibinfo{title}{{Ska Sort}}.
\newblock \bibinfo{howpublished}{\url{https://github.com/skarupke/ska_sort}}.
\newblock


\bibitem[Thiel(2019)]%
        {Thiel_2019}
\bibfield{author}{\bibinfo{person}{Stuart Thiel}.}
  \bibinfo{year}{2019}\natexlab{}.
\newblock \bibinfo{title}{Diverting Fast Radix}.
\newblock
\newblock
\urldef\tempurl%
\url{https://github.com/ramou/dfr}
\showURL{%
\tempurl}


\bibitem[Thiel et~al\mbox{.}(2016)]%
        {ThielImprovingGraphChi2016}
\bibfield{author}{\bibinfo{person}{Stuart Thiel}, \bibinfo{person}{Greg
  Butler}, {and} \bibinfo{person}{Larry Thiel}.}
  \bibinfo{year}{2016}\natexlab{}.
\newblock \showarticletitle{{Improving GraphChi for Large Graph Processing:
  Fast Radix Sort in Pre-Processing}}. In
  \bibinfo{booktitle}{\emph{{P}roceedings of the 20th {I}nternational
  {D}atabase {E}ngineering \& {A}pplications {S}ymposium}} (Montreal, QC,
  Canada) \emph{(\bibinfo{series}{IDEAS '16})}. \bibinfo{publisher}{Association
  for Computing Machinery}, \bibinfo{address}{New York, NY, USA},
  \bibinfo{pages}{135--141}.
\newblock
\showISBNx{978-1-4503-4118-9}
\urldef\tempurl%
\url{https://doi.org/10.1145/2938503.2938554}
\showDOI{\tempurl}


\end{thebibliography}

\end{document}